\documentclass[10pt,conference]{IEEEtran}
\usepackage{epsf,psfrag,amssymb,amsfonts,cite,amsmath,eufrak,yfonts}
\usepackage{graphicx}
\newtheorem{theorem}{Theorem}
\newtheorem{example}{Example}

\newtheorem{lemma}{Lemma}
\newtheorem{definition}{Definition}

\def\psfancypar#1#2{\begingroup\def\par{\endgraf\endgroup\lineskiplimit=0pt}
               \setbox2=\hbox{\large\sc #2}
               \newdimen\tmpht \tmpht \ht2 \advance\tmpht by \baselineskip
               \font\hhuge=Times-Bold at \tmpht
               \setbox1=\hbox{{\hhuge #1}}
               \count7=\tmpht \count8=\ht1
               \divide\count8 by 1000 \divide\count7 by \count8
               \tmpht=.001\tmpht\multiply\tmpht by \count7
               \font\hhuge=Times-Bold at \tmpht
               \setbox1=\hbox{{\hhuge #1}}
               \noindent
                \hangindent1.05\wd1
               \hangafter=-2 {\hskip-\hangindent
               \lower1\ht1\hbox{\raise1.0\ht2\copy1}%
                \kern-0\wd1}\copy2\lineskiplimit=-1000pt}

\newcommand{\beq}{\begin{equation}}
\newcommand{\eeq}{\end{equation}}
\newcommand{\bqa}{\begin{eqnarray}}
\newcommand{\eqa}{\end{eqnarray}}
\newcommand{\bqn}{\begin{eqnarray*}}
\newcommand{\eqn}{\end{eqnarray*}}
\newcommand{\nn}{\nonumber}

\newcommand{\be}{\begin{enumerate}}
\newcommand{\ee}{\end{enumerate}}
\newcommand{\bi}{\begin{itemize}}
\newcommand{\ei}{\end{itemize}}
\newcommand{\bd}{\begin{description}}
\newcommand{\ed}{\end{description}}
\newcommand{\ba}{\begin{array}}
\newcommand{\ea}{\end{array}}
\newcommand{\bde}{\begin{definition}}
\newcommand{\ede}{\end{definition}}
\newcommand{\bex}{\begin{example}}
\newcommand{\eex}{\end{example}}

\def\boxit#1{\vbox{\hrule\hbox{\vrule\kern3pt
        \vbox{\kern3pt#1\kern3pt}\kern3pt\vrule}\hrule}}

\def\reals{ { {\rm  I \kern-0.15em R }  } }
\def\complex{ {\,{{\rm C} \kern-0.50em \raise0.20ex {  |}}\, }}

\def\0bf{{\bf 0}}
\def\1bf{{\bf 1}}
\def\2bf{{\bf 2}}
\def\3bf{{\bf 3}}
\def\4bf{{\bf 4}}
\def\5bf{{\bf 5}}
\def\6bf{{\bf 6}}
\def\7bf{{\bf 7}}
\def\8bf{{\bf 8}}
\def\9bf{{\bf 9}}

\def\Rbf{{\bf R}}

\def\Xbf{{\bf X}}

\def\Mmat{\mathcal{M}}

\def\Wmat{\mathcal{W}}
\def\Xmat{\mathcal{X}}
\def\Ymat{\mathcal{Y}}

\def\Rxx{\Rbf_{\ssstyle X\kern-.1em X}}

\let\ssstyle=\scriptscriptstyle

\def\Kout{\setbox1=\hbox{\Huge\bf K}\hbox to
1.05\wd1{\hspace{.05\wd1}
}}
\def\Sout{\setbox1=\hbox{\Huge\bf S}\hbox to 1.05\wd1{\hspace{.05\wd1}
}}

\def\scalefig#1{\epsfxsize #1\textwidth}
\begin{document}
\title{\bf \LARGE The Common Information of $N$ Dependent Random Variables }
\IEEEoverridecommandlockouts

\author{\authorblockN{Wei Liu}
\authorblockA{
Department of EECS\\
Syracuse University\\
 Email: wliu28@syr.edu} \and
 \authorblockN{Ge Xu}
\authorblockA{
Department of EECS\\
Syracuse University\\
 Email: gexu@syr.edu} \and
 \authorblockN{Biao Chen}
\authorblockA{
Department of EECS\\
Syracuse University\\
 Email: bichen@syr.edu}}

\maketitle
\begin{abstract} This paper generalizes Wyner's definition of common information of a pair of random variables to that of
 $N$ random variables. We prove coding theorems that show the same operational meanings for the common information of two
 random variables generalize to that of $N$ random variables. As a byproduct of our proof, we show that the Gray-Wyner source
 coding network can be generalized to $N$ source squences with $N$ decoders. We also establish a monotone property of Wyner's common information
 which is in contrast to other notions of the common information, specifically Shannon's mutual information and
 G\'{a}cs and K\"{o}rner's common randomness. Examples about the computation of Wyner's common information of $N$ random
 variables are also given.
\end{abstract}
\section{Introduction}
Consider  a pair of dependent random variables $X$ and $Y$ with
joint distribution $P(x,y)$. Characterizing the common information
between $X$ and $Y$ has been a topic of research interest in the
past
decades\cite{Gacs&korner,Ahlswede&korner,Witsenhausen:1975,Wyner_CI_75IT,Witsenhausen:1976}.
There have been three classical notions reported in the
literature. 

\noindent {\bf Shannon's \cite{Shannon:1948} mutual information
$I(X;Y)$}

Shannon's mutual information measures how much uncertainty can be
reduced with respect to one random variable by observation the
other random variable. In the case that $X$ and $Y$ are
independent, mutual information $I(X;Y)=0$, indicating that
observing one variable $X$ does not give any information about $Y$
and vice versa.
Shannon's mutual information carries operational meanings that are
instrumental in laying the foundation for information theory.

\noindent {\bf G\'{a}cs and K\"{o}rner's \cite{Gacs&korner} common
randomness $K(X,Y)$}

Consider a pair of independent and identically distributed random
sequences $X^n,Y^n$ with each pair $(X_i,Y_i)\sim P(x,y)$. These
two sequences are observed respectively by two nodes, which
attempt to map the sequences onto a common message set $\Wmat$.
Specifically, let $f_n$ and $g_n$ be such mappings, i.e., \bqn
f_n: && \Xmat^n \rightarrow \Wmat, \\
g_n: && \Ymat^n \rightarrow \Wmat. \eqn Define
$\epsilon_n=Pr(W_1\neq W_2)$ where $W_1=f_n(X^n)$ and
$W_2=g_n(Y^n)$. G\'{a}cs and K\"{o}rner's common randomness is
defined as
\[
K(X,Y)=\lim_{n\rightarrow\infty,\epsilon_n\rightarrow 0} \sup
\frac{1}{n}H(W_1).
\]
G\'{a}cs and K\"{o}rner's  common randomness has found extensive applications in cryptography, i.e., for key generation \cite{Maurer,Ahlswede&csiszar1,Ahlswede&csiszar2}.
On the other hand, the common randomness notion is rather
restrictive as it equals $0$ in most cases except for the
following special case (or random variable pairs that can be
converted to such distributions through relabeling of
realizations, i.e., permutation of joint distribution matrix). Let
$X$ and $Y$ be $X=(X',V)$ and $Y=(Y',V)$, respectively, where
$X',Y',V$ are independent. Clearly, the common part between $X$
and $Y$ is $V$ and it follows that $K(X;Y)=H(V)$. Note that for
this example $I(X;Y)=K(X;Y)=H(V)$.

\noindent {\bf Wyner's \cite{Wyner_CI_75IT} common information
$C(X,Y)$}

\begin{figure}
\centerline{
\begin{psfrags}
\psfrag{xnyn}[c]{$X^n, Y^n$}
\psfrag{encoder}[c]{Encoder}
\psfrag{decoder1}[c]{Decoder 1}
\psfrag{decoder2}[c]{Decoder 2}
\psfrag{w1}[c]{$W_1$}
\psfrag{w2}[c]{$W_2$}
\psfrag{w0}[c]{$W_0$}
\psfrag{xnh}[c]{$\hat{X}^n$}
\psfrag{ynh}[c]{$\hat{Y}^n$}
 \scalefig{.45}\epsfbox{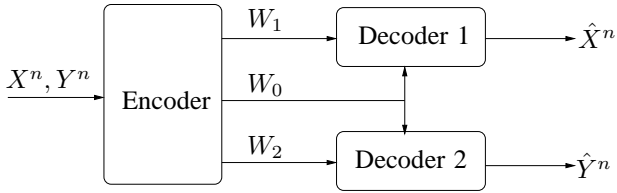}
\end{psfrags}}
\caption{\label{fig:model1}Source coding over a simple network.}
\end{figure}

Wyner's common information is defined as
 \beq
C(X,Y)=\min_{X\rightarrow W\rightarrow Y} I(XY;W). \eeq Thus the
hidden (or auxiliary) variable $W$ induces a Markov chain $X-W-Y$,
or, equivalently, a conditional independence structure of $X,Y$
being independent given $W$.
 Wyner gave two operational meanings for the above definition.
 The first approach is shown in Fig. \ref{fig:model1}. The encoder observes a pair
 of sequences $(X^n,Y^n)$, and map
 them to three messages $W_0,W_1,W_2$, taking values in alphabets of respective sizes
 $2^{nR_0},2^{nR_1},2^{nR_2}$.
 Decoder 1, upon receiving $(W_0,W_1)$, needs to reproduce $X^n$ reliably
 while decoder 2, upon receiving $(W_0,W_2)$, needs to reproduce $Y^n$ reliably.
 Let $C_1$ be the infimum of all admissible $R_0$ for the system in Fig. 1 such that
 the total rate  $R_0+R_1+R_2\approx H(X,Y)$.

The second approach is shown in Fig. 2. A common input $W$,
uniformly distributed on ${\cal W}=\{1,\cdots,2^{nR_0}\}$ is given
to two separate processors which are otherwise independent of
each other. These processors (random variable generators)
generating independent and identically distributed sequences
according to $q_1(X|W)$ and $q_2(Y|W)$ respectively. The output
sequences of the two processors are denoted by ${\tilde X}^n$ and
${\tilde Y}^n$ respectively. Thus the joint distribution of the
output sequences is, \beq Q({\tilde X}^n,{\tilde
Y}^n)=\sum_{w\in{\cal W}}\frac{1}{{\cal
W}}q_1(X^n|W)q_2(Y^n|W).\label{eq:sta}\eeq Define $C_2$ of $(X,Y)$
to be infimum of rate $R_0$ for the common input such that
$q({\tilde X}^n,{\tilde Y}^n)$ close to  $p(X^n,Y^n)$, where the
closeness is defined using the average divergence of the two
distributions
 \bqa D_n(P,Q)=\frac{1}{n} \sum_{x^n\in{\cal
X}^n,y^n\in {\cal
Y}^n}P(x^n,y^n)\log{\frac{P(x^n,y^n)}{Q(x^n,y^n)}}.\eqa

Wyner proved that
 \beq
C_1=C_2= C(X,Y).
\eeq

It was observed in \cite{Wyner_CI_75IT} that \beq K(X,Y)\leq
I(X;Y)\leq C(X,Y).\eeq Wyner \cite{Wyner_CI_75IT} and
Witsenhausen\cite{Witsenhausen:1976} also provide several examples
on how to calculate the common information $C(X,Y)$. For the
example of $X=(X',V)$ and $Y=(Y',V)$ with $(X',Y',V)$ mutually
independent, $C(X,Y)=I(X;Y)=K(X,Y)=H(V)$.

Generalizing of mutual information to $N$ random variables  was
first reported in \cite{Hu62}. The generalization comes from the
observation that for a pair of random variables, Shannon's
information measures is consistent with the Venn diagram for set
operation and a comprehensive treatment was available in
\cite{CK:book,Yeung:Book08}. G\'{a}cs and K\"{o}rner's
common randomness was recently generalized to multiple random
variables by Tyagi, Narayan and Gupta  in
\cite{DBLP:journals/corr/abs-1007-2945}, which extends the
encoding process in the definition of common randomness to that of
$N$ terminals.

In this paper, we generalize Wyner's common information of a pair
of random variables to that of $N$ dependent variables. We show
that the operational meaning defined in both approaches  are
still valid. Moreover, we establish some monotone property of such
generalization which contrast to the notion of `common'
information. Specifically, we show that the common information
does {\em not} decrease as the number of variables increases while
keeping the same marginal distribution. This is different from the
other two notions of common information. Examples on evaluating
$C(X_1,X_2,\cdots,X_N)$ are given for circularly symmetric binary
sources and the asymptotic results are also studied.

The rest of this paper is organized as follows. Section II gives
the problem formulation and main results. Section III gives some
examples and discussions.  Section IV concludes the paper.

\begin{figure}
\centerline{
\begin{psfrags}
\psfrag{W}[r]{$W$}
\psfrag{PROCESSOR 1}[l]{Processor 1}
\psfrag{PROCESSOR 2}[l]{Processor 2}
\psfrag{Xn}[c]{${\tilde X}^n$}
\psfrag{Yn}[c]{${\tilde Y}^n$}
 \scalefig{.3}\epsfbox{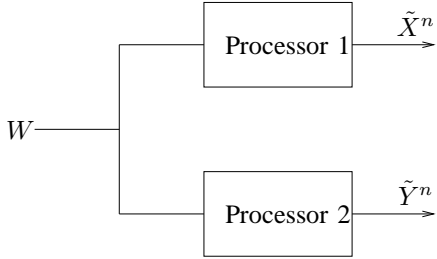}
\end{psfrags}}
\caption{\label{fig:model2}Random variable generators.}
\end{figure}
\section{Problem Statement and Main Results}
 Let $X_1,X_2,\cdots,X_N$ be random variables that take values on the finite alphabet sets
 ${\cal X}_1, {\cal X}_2, \cdots, {\cal X}_N$ with joint distribution
 $P(x_1,x_2,\cdots,x_N)$. Our generalization of Wyner's common
 information is to define a similar measure for $N$ random
 variables by preserving the conditional independence structure through the introduction of
 an auxiliary random variable.
Specifically, we define \beq C(X_1,X_2,\cdots,X_N)\triangleq \inf
I(X_1,X_2,\cdots,X_N;W),\eeq where the infimum is taken over all
the joint distributions of $(X_1,X_2,\cdots,X_N,W)$ such that \bqa
\sum_{w}P(x_1,x_2,\cdots,x_n,w)&=&P(x_1,x_2,\cdots,x_N),
\label{theorem1.1}\\
P(x_1,...,x_n|w)&=&\prod_{i=1}^{n}P(x_i|w).\label{theorem1.2} \eqa
Thus the marginal distribution of $(X_1,X_2,\cdots,X_N)$ is
$P(x_1,x_2,\cdots,x_N)$ and $(X_1,\cdots,X_N)$ are conditionally
independent given $W$.

We now give formal definitions of $C_1$ and $C_2$ for $N$ random
variables.
Consider $N$ length-$n$ independent and identically distributed
source sequences $(x^n_1,x^n_2,\cdots,x^n_N)$ with
$(X_{1i},X_{2i},\cdots,X_{Ni})\sim p(x_1,x_2,\cdots,x_N)$, i.e.,
\beq P^{(n)}(x^n_1,x^n_2,\cdots,x^n_N)=\prod^n_{i=1}
P(x_{1i},x_{2i},\cdots,x_{Ni}).\label{eq:pn}\eeq


For the Gray-Wyner source coding network, we start with the
definition of encoder-decoders.
\begin{definition}
A $(n,{\cal M}_0,{\cal M}_1,\cdots,{\cal M}_N)$ code consists of
the following: \begin{itemize} \item An encoder mapping
\[
 f:  {\cal X}^n_1 \times {\cal X}^n_2 \times \cdots \times {\cal X}^n_N \rightarrow  \Mmat_0\times \Mmat_1 \times \cdots \times
 \Mmat_N,
 \]
where $\Mmat_i = \{1,2,\cdots,2^{nR_i}\}$.

\item $N$ decoders $g_i$,  for $i=1,2,\cdots,N$,
 \beq g_i:
\Mmat_i\times \Mmat_0 \rightarrow {\cal X}^n_i. \eeq
\end{itemize}
\end{definition}

The probability of error is defined as \beq
P^{(n)}_e=Pr\{(\hat{X}^n_1\hat{X}^n_2\cdots\hat{X}^n_N)\neq
(X^n_1,X^n_2,\cdots,X^n_N)\},\eeq where $\hat{X}^n_i=g_i(M_i,M_0)$
for $i=1,\cdots, N$.

\begin{definition}\label{def2}
A number $R_0$ is said to be {\em achievable} if for any
$\epsilon>0$, we can find an $n$ sufficiently large such that
there exists a  $(n,{\cal M}_0,{\cal M}_1,\cdots,{\cal M}_N)$ code
with
\bqa {\cal M}_0&\leq& 2^{nR_0}\label{eq:achi0}\\
P^{(n)}_e&\leq& \epsilon ,\label{eq:achi1}\\
     \frac{1}{n}\sum^N_{i=0}\log{{\cal M}_i}&\leq&H(X_1,X_2,\cdots,X_N)+\epsilon.\label{eq:ach2}\eqa
\end{definition}
As with the case for two random variables, $C_1$ is defined as the
infimum of all achievable $R_0$.

For the second approach of approximating joint distribution, we
again start with the following definition.
\begin{definition}
An $(n, {\cal M}, \Delta) $ {\em generator} consists of the
following: \bi \item a message set ${\cal W}\in \{1,2,\cdots,
2^{nR}\}$; \item for all $w\in {\cal W}$ and $N$ conditional
probability distributions $q^{(n)}_i(x^n_i|w)$, for
$i=1,2,\cdots,N$, define the probability distribution on $ {\cal
X}^n_1 \times {\cal X}^n_2 \times \cdots \times {\cal X}^n_N$ \beq
Q^{(n)}(X^n_1,X^n_2,\cdots,X^n_N)=\sum_{w\in{\cal
W}}\frac{1}{{\cal M}}\prod^N_{i=1}
q^{(n)}_i(x^n_i|w).\label{eq:qn}\eeq \ei
\end{definition}
Thus the $N$ processors serve as random number generators each
generating independent and identically distributed (i.i.d.) sequence $\hat{X}_i^n$ according to $q(x_i|w)$
and the output of the processors follow joint distribution defined
in (\ref{eq:pn}).
Let \beq
\Delta=D_n(P^{(n)};Q^{(n)})=\frac{1}{n}\sum_{x^n_i\in{X}^n_i,i=1,2,\cdots,N}P^{(n)}
\log{\frac{P^{(n)}}{Q^{(n)}}},
\eeq where $P^{(n)}$ and $Q^{(n)}$
are defined as in (\ref{eq:pn}) and (\ref{eq:qn}) respectively.

\begin{definition}
A number $R$ is said to be achievable if for all $\epsilon>0$, we can find an $n$ sufficiently large
such that there exists a $(n,{\cal M},\Delta)$ generator with ${\cal M}\leq 2^{nR}$ and $\Delta \leq \epsilon$.
\end{definition}
We  define $C_2$ as the infimum of all achievable $R$.

The main result of this paper is the following theorm.

\begin{theorem}
\beq C_1=C_2=C(X_1,X_2,\cdots,X_N).\eeq
\end{theorem}
The proof of Theorem 1 is given in the Appendix. Thus both $C_1$
and $C_2$ admit single letter characterization which coincides
with $C(X_1,\cdots,X_N)$.

\section{Examples and discussions}

We start with the following example. Let $X=(X^{'},U,V)$,
$Y=(Y^{'},V,W)$ and $Z=(Z^{'},W,U)$ where the random variables
$X^{'},Y^{'},Z^{'},U,V,W$ are mutually independent. It is easy to
show that for this example
\[
I(X;Y;Z)=K(X,Y,Z)=0,
\]
whereas
\[
C(X,Y,Z)=H(UVW).
\]
On the other hand, \bqn C(X,Y)&=&H(V),\\
C(X,Z)&=&H(U),\\
C(Y,Z)&=&H(W). \eqn

What is interesting is that the inclusion of an additional
variable increases the common information. This is somewhat
surprising: if the information is {\em common} it ought to be
non-increasing when more random variables are included. Indeed, we
can prove the following general result:

\begin{lemma}
Let $(X_1,\cdots,X_N)\sim p(x_1,\cdots,x_N)$. For any two sets
$A,B$ that satisfy $A\subseteq B \subseteq {\cal
N}=\{1,2,\cdots,N\}$,
 \beq
 C(\Xbf_{A})\leq C(\Xbf_{B}),
\eeq where $\Xbf_A=\{X_i,i\in A\}$ and $\Xbf_B=\{X_i,i\in B\}$.
\end{lemma}
\proof  Let $W'$ be the $W$ that achieves $C(\Xbf_{B})$, i.e.,
$I(W';\Xbf_B)=\inf I(W;\Xbf_B)$. But $A\subseteq B$, thus $\Xbf_B$
conditionally independent given $W'$ implies that $\Xbf_A$ is
conditionally independent given $W'$. Thus \bqn I(\Xbf_B;W')&\geq&
I(\Xbf_A;W')\\
&\geq & \inf I(\Xbf_A;W) \eqn where the infimum is taken over all
$W$ such that $\Xbf_A$ is independent given $W$.

This monotone property perhaps suggests that the name common
information, while meaningful for pair of variables, no longer
suits the generalization to $N$ variables. We comment here that
G\'{a}cs and K\"{o}rner's common randomness follows a different
monotone property
\[
 K(\Xbf_{A})\geq K(\Xbf_{B})
 \]
 while there is no definitive inequality relationship for mutual
 information.

As a consequence, we have for any $N$ random variables
\[
 C(X_1,X_2,\cdots,X_N)\geq K(X_1,X_2,\cdots,X_N).
 \]

We now examine another example in which Wyner's common information
increases as the number of the observations increases. Moreover
the common information eventually converges and the asymptote
suggests that the notion of common information may have potential
application in certain inference problem.

 Consider first the example of three binary random variables $X_1, X_2,X_3$ with
 joint distribution

 \bqa P(x_1,x_2,x_3)=\left\{\begin{array}{ll}
\frac{1}{2}-\frac{3}{4}a_0 &\mbox{if $x_1=x_2=x_3$}\\
\frac{1}{4}a_0 &\mbox{otherwise}
\end{array} \right. \label{eq:joint}\eqa
where the parameter $a_0$ satisfies $0\leq a_0\leq \frac{1}{2}$.

It can be easily verified that
\beq Pr\{X_i=0\}=\frac{1}{2},\eeq
 for $i=1,2,3$ and that for $1\leq i,j\leq 3, i\neq j$,
\beq Pr(X_i=x_i,X_j=x_j)=\frac{1}{2}(1-a_0)\delta_{x_i,x_j}+\frac{1}{2}a_0(1-\delta_{x_i,x_j}),\label{eq:3source}\eeq
where $\delta_{a,b}=1$ if $a=b$ and $0$ otherwise.

Thus, each pair of $(X_i,X_j)$, $i \neq j$, can be viewed as a
doubly symmetric binary source  as defined in
\cite{Wyner_CI_75IT}. We refer to this set of exchangeable binary
sources circularly symmetric binary source. For such circularly
symmetric binary source $(X_1,X_2,X_3)$ with joint distribution
given in (\ref{eq:joint}) and random variables $(X_1,X_2,X_3,W)$
that satisfy (\ref{theorem1.1}) and (\ref{theorem1.2}), we have
the following lemma.

\begin{figure}
\centerline{
\begin{psfrags}
\psfrag{w}[c]{$\quad W$}
\psfrag{x1}[c]{$\quad{X}_1$}
\psfrag{x2}[c]{$\quad{X}_2$}
\psfrag{xn}[c]{$\quad{X}_N$}
\psfrag{dot}[c]{$\cdots$}
 \scalefig{.3}\epsfbox{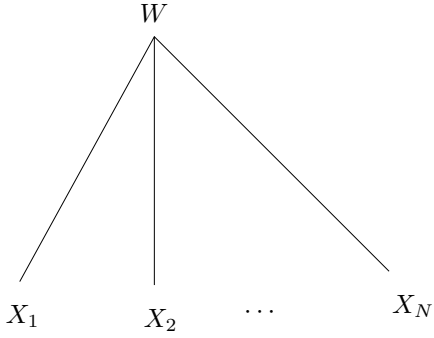}
\end{psfrags}}
\caption{\label{fig:model3} A simple Bayesian graph model.}
\end{figure}

\begin{lemma}\label{lemma2}
\beq H(X_1|W)+H(X_2|W)+H(X_3|W)\leq 3h(a_1),\eeq
where $a_1=\frac{1}{2}-\frac{1}{2}(1-2a_0)^{\frac{1}{2}}$.
\end{lemma}
This lemma is a direct consequence of Wyner's result on doubly symmetric binary source \cite{Wyner_CI_75IT}. Therefore, we have,
\bqa &&I(X_1X_2X_3;W)\nn\\
 &=& H(X_1X_2X_3)-H(X_1X_2X_3|W),\nn\\
                      &=& H(X_1X_2X_3)-\sum^3_{i=1}H(X_i|W),\nn\\
                      &\geq& H(X_1X_2X_3)-3h(a_1),\nn\\
                       &=& 1+h(a_0)+a_0+(1-a_0)h\left(\frac{a_0}{2(1-a_0)}\right)-2h(a_1),\label{eq:Ilower}\nn\\
                       && \eqa
This lower bound can indeed be  achieved by choosing the following
random variables. Let $W$  be a random variable with
$p_W(0)=p_W(1)=1/2$, i.e., a Bernoulli$(1/2)$ random variable. Let
 each $X_i$ be the output of a binary symmetric channel (BSC) with crossover probability
 $a_1$ with $W$ as input. The channels share the common input $W$
 but are otherwise independent of each other. This is illustrated
 in the simple Bayesian graph model  in Fig. \ref{fig:model3}
with $N=3$ where each link represents a BSC with crossover
probability $a_1$.

Thus, the common information of this circularly symmetric binary
source is,
\bqa C(X_1,X_2,X_3)&=&1+a_0+h(a_0)+\nn\\
&&(1-a_0)h\left(\frac{a_0}{2(1-a_0)}\right)-3h(a_1),\nn\\
&&\eqa

\begin{figure}[tb]
\centerline{\leavevmode \epsfxsize=2.85in \epsfysize=2.35in
\epsfbox{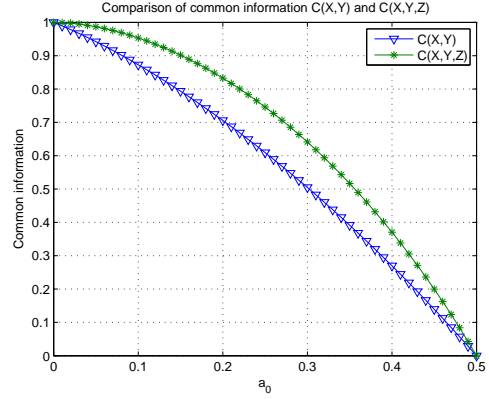}}
\caption{\label{fig:comp} Comparison of common information.}
\end{figure}

Notice that any pair of $(X_i,X_j)$ is a doubly symmetric binary
source \cite{Wyner_CI_75IT}, therefore,
\[
C(X,Y)=1+h(a_0)-2h(a_1).
\]
It is straightforward to check that
\[
C(X,Y,Z)>C(X,Y) \] when $0<a_0<\frac{1}{2}$. This is also shown
numerically in Fig. \ref{fig:comp}.

We now study the generalization of above example to arbitrary $N$
and in particular the asymptotic value of the common information
for the circularly symmetric binary sources.

Consider $N$ binary random variables $X_1,X_2,\cdots,X_N$ with
joint distribution $p(X_1,X_2,\cdots,X_N)$ generated by an
underlying Bayesian graph model as in Fig. \ref{fig:model3}, where $W$ is
a Bernoulli$(1/2)$ random variable and each $X_i$,
$i=1,2,\cdots,N$, is the output of a BSC with crossover
probability $a_1 (0\leq a_1\leq\frac{1}{2}) $ with a common input
$W$. Hence, for $x_1,x_2,\cdots,x_N \in\{0,1\}$,
    \beq P(x_1,x_2,\cdots,x_n)=\sum_{w\in\{0,1\}}\frac{1}{ 2}\prod^N_{i=1} P_i(x_i|w),\label{eq:jointN}\eeq
    where for each $i=1,2,\cdots,N$, $p_i(x_i|w)=(1-a_1)$ if $x_i=w$ and $a_1$ otherwise.

Similarly, we have,    \beq \sum^{N}_{i=1}H(X_i|W)\leq N h(a_1),\eeq
    for any random variable $W$ that satisfies (\ref{theorem1.1}) and (\ref{theorem1.2}).

Therefore, $C(X_1,X_2,\cdots,X_N)$ can be lower bounded by
    \beq C(X_1,X_2,\cdots,X_N)\geq H(X_1,X_2,\cdots,X_N)-Nh(a_1).\eeq

    On the other hand, the above lower bound is achievable by exactly the same $W$ in the above
    Bayesian model. Hence, we have,
    \beq C(X_1,X_2,\cdots,X_N)=H(X_1,X_2,\cdots,X_N)-Nh(a_1),\eeq
    where $H(X_1,X_2,\cdots,X_N)$ can be calculated from (\ref{eq:jointN}).

Now consider the above model but with increasing $N$. For any
$\epsilon$ and $a_1<1/2$, it is clear that
\[
H(W|X_1,X_2,\cdots,X_N)<\epsilon
\]
for $N$ sufficiently large. This can be established by the Fano's
inequality as one can estimate $W$ with arbitrary reliability
given $X_1,\cdots,X_N$ for sufficiently large $N$. Therefore,
    \bqa &&C(X_1,X_2,\cdots,X_N)\nn\\
    &=&H(X_1,X_2,\cdots,X_N)-Nh(a_1),\nn\\
    &=&H(X_1,X_2,\cdots,X_N,{W})-Nh(a_1)\nn\\
    &&-H({W}|X_1,X_2,\cdots,X_N),\nn\\
    &\geq&H(W)-\epsilon,\eqa
    where the last step is from the fact that
    $H(X_1,X_2,\cdots,X_N|{W})=Nh(a_1)$.
On the other hand,
\[
C(X_1,\cdots,X_N)\leq H(W)
\]
for any $N$. Thus, for $a_1<1/2$, \[ \lim_{N\rightarrow
\infty}C(X_1,X_2,\cdots,X_N) =H(W)=1
\]
If $a_1=1/2$, then $X_1,\cdots,X_N$ are mutually independent hence
$C(X_1,\cdots,X_N)=0$.

\section{Conclusions}
This paper generalized Wyner's common information, defined for a
pair of random variables, to that of $N$ dependent random
variables. We showed that it is the minimum common information
rate $R_0$ needed for $N$ separate decoders to recover their
intended sources losslessly while keeping the total rate close to
the entropy bound. It is also equivalently to the smallest rate of
the common input to $N$ independent processors (random number
generators), such that the output distribution is approximately
the same as the given joint distribution. It was shown that such
generalization leads to the phenomenon of `common' information
non-decreasing as the number of sources increases.

For the example of circularly symmetric binary sources, we show
that common information not only increases as $N$ grows, but
eventually converges to the entropy of $W$ that achieves
$C(X_1,\cdots,X_N)$.

\appendix
In this appendix, we give the proof of Theorem 1. First, as with
\cite{Wyner_CI_75IT}, we define a quantity
$\Gamma(\delta_1,\delta_2)$ which plays an important role in the
proof.

Let $(X_1,X_2,\cdots,X_N)\sim P(x_1,x_2,\cdots,x_N)$ where
$X_1,\cdots,X_N$ take values in finite alphabet
$\Xmat_1,\cdots,\Xmat_N$. Let
$(\hat{X}_1,\hat{X}_2,\cdots,\hat{X}_N,W)$ be a $(N+1)$tuple of
random variables where $\hat{X}_1\in {\cal X}_1, \hat{X}_2\in{\cal
X}_2, \cdots, \hat{X}_N\in{\cal X}_N$ and $W\in {\cal W}$, a
finite set. Denote the marginal distribution of
$(\hat{X}_1,\hat{X}_2,\cdots,\hat{X}_N)$ by \beq
Q(x_1,x_2,\cdots,x_N)=Pr(\hat{X}_1=x_1,\hat{X}_2=x_2,\cdots,\hat{X}_N=x_n),\eeq
for $x_i\in {\cal X}_i$, $i=1,2,\cdots,N$.

For any $\delta_1,\delta_2 \geq 0$, define
\beq \Gamma(\delta_1,\delta_2)= \sup H(\hat{X}_1,\hat{X}_2,\cdots,\hat{X}_N|W), \eeq
where the sumpremum is taken over all $(N+1)$-tuples $(\hat{X}_1,\hat{X}_2,\cdots,\hat{X}_N,W)$ that satisfy
\beq D(P;Q)=\sum_{x,y}P(x_1,x_2,\cdots,x_N)\log{\frac{P(x_1,x_2,\cdots,x_N)}{Q(x_1,x_2,\cdots,x_N)}}\leq \delta_1,\label{eq:gamma1}\eeq
and \beq \sum^N_{i=1}H(\hat{X}_i|W)-H(\hat{X}_1,\hat{X}_2,\cdots,\hat{X}_N|W)\leq \delta_2.\label{eq:gamma2} \eeq
It follows that $C(X_1,X_2,\cdots,X_N)$ as defined in Theorem 1, is equivalent to
\beq C(X_1,X_2,\cdots,X_N)=H(X_1,X_2,\cdots,X_N)-\Gamma(0,0).\eeq

The following lemma gives some properties of $\Gamma(\delta_1,\delta_2)$.
\begin{lemma} \label{lemmawnyer}

1) For all $\delta_1,\delta_2\geq 0$, there exists a $(N+1)$-tuple $(\hat{X}_1,\hat{X}_2,\cdots,\hat{X}_N,W)$ such that (\ref{eq:gamma1}) and (\ref{eq:gamma2}) are satisfied and
\beq \Gamma(\delta_1,\delta_2)=H(\hat{X}_1,\hat{X}_2,\cdots,\hat{X}_N|W).\eeq
Moreover, for $\delta_1,\delta_2=0$,
\beq |{\cal W}|\leq \prod^N_{i=1}|{\cal X}_i|.\eeq

2) $\Gamma(\delta_1,\delta_2)$ is a concave function of $(\delta_1,\delta_2)$ and it is continuous for all $\delta_1,\delta_2\geq0$.

3) For $\delta\geq0$, define $\Gamma_1(\delta)=\Gamma(0,\delta)$ and $\Gamma_2(\delta)=\Gamma(\delta,0)$, then
$\Gamma_1(\delta)$ and $\Gamma_2(\delta)$ are concave and continuous for $\delta\geq 0$.
\end{lemma}

The proof of Lemma 1 follows similarly as the proof of Theorem
$4.4$ in \cite{Wyner_CI_75IT}.

\subsection{Proof of $C_1=C(X_1,X_2,\cdots,X_N).$}
In this section, we prove the first part of Theorem 1, that is
$C_1=C(X_1,X_2,\cdots,X_N).$ We first prove the converse part,
that is for any $R_0$ that is achievable for the Gray-Wyner source
coding network, we have,
\begin{theorem}[Converse]
\beq C_1\geq C(X_1,X_2,\cdots,X_N).\eeq
\end{theorem}

To prove the converse, first let $(f,g_i)$, $i=1,2,\cdots,N$ be any $(n,{\cal M}_0,{\cal M}_1,\cdots,{\cal M}_N)$ code that satisfies (\ref{eq:achi0}), (\ref{eq:achi1}) and (\ref{eq:ach2}).

 Then, we have,
\bqa &&\log{{\cal M}_0}\nn\\
&\geq& H(M_0),\\
&\geq& I(X^n_1X^n_2\cdots X^n_N;M_0),\\
&=&H(X^n_1X^n_2\cdots X^n_N)-H(X^n_1X^n_2\cdots X^n_N|M_0),\\
&=&nH(X_1X_2\cdots X_N)-\sum^n_{j=1}H(X_{1j}X_{2j}\cdots X_{Nj}|W_j),\label{eq:r0temp}\eqa
where $W_j\triangleq(M_0,X^{j-1}_1,X^{j-1}_2,\cdots,X^{j-1}_N)$ and $X^{j-1}_i=(X_{i1},X_{i2},\cdots,X_{i,j-1})$ for $i=1,2,\cdots,N$.

Notice that, the $(N+1)$-tuple $(X_{1j},X_{2j},\cdots,X_{Nj},W_j)$ satisfies condition (\ref{eq:gamma1}) and (\ref{eq:gamma2}) with $\delta_1=0$ and
\beq \delta^{(j)}_2=\sum^N_{i=1}H(X_{i,j}|W_j)-H(X_{1j},X_{2j},\cdots,X_{Nj}|W_j).\eeq

Hence, by the definition of $\Gamma(\delta_1,\delta_2)$, we have
\beq H(X_{1j}X_{2j}\cdots X_{Nj}|W_j)\leq \Gamma_1(\delta^{(j)}_2).\label{eq:delta_2j}\eeq
Substitute  (\ref{eq:delta_2j}) into (\ref{eq:r0temp}), we get,
\bqa \log{\cal M}_0&\geq& nH(X_1X_2\cdots X_N)-\sum^n_{j=1}\Gamma_1(\delta^{(j)}_2),\\
&\geq&nH(X_1X_2\cdots X_N)-n\Gamma_1(\frac{1}{n}\sum^n_{j=1}\delta^{(j)}_2). \label{eq:r0bound}\eqa
where the last step is from the concavity of $\Gamma_1(\cdot)$ function. Now define
\beq \eta=\frac{1}{n}\sum^n_{j=1}\delta^{(j)}_2.\eeq
The following lemma gives an upper bound on $\eta$.

\begin{lemma}\label{lemma2}
 For any $(n,{\cal M}_0,{\cal M}_1,\cdots,{\cal M}_N)$ code that satisfies (\ref{eq:achi0}),  (\ref{eq:achi1}) and (\ref{eq:ach2}), we have
 \beq \eta\leq (N+1)\epsilon.\eeq
\end{lemma}

{\em Proof} :

By Fano's inequality, we have, for $i=1,2,\cdots,N$,
\beq H(X^n_i|M_0M_i)\leq n\epsilon.\eeq
 Hence, we have, for $i=1,2,\cdots,N$,
\bqa \log{{\cal M}_i}&\geq& H(M_i),\\
&\geq& H(M_i|M_0),\\
&=&H(X^n_iM_i|M_0)-H(X^n_i|M_iM_0),\\
&\geq&H(X^n_iM_i|M_0)-n\epsilon,\\
&=&H(X^n|M_0)-n \epsilon.\eqa
Then, we get,
\beq \sum^N_{i=1}\log{{\cal M}_i}\geq\sum^N_{i=1}H(X^n_i|M_0)-n\epsilon'.\eeq
where $\epsilon'=N\epsilon$. Together with (\ref{eq:r0temp}), we get,
\bqa
\sum^N_{i=0}\log{{\cal M}_i}&\geq& nH(X_1X_2\cdots X_N)\nn\\
&&-\sum^n_{j=1}H(X_{1j}X_{2j}\cdots X_{Nj}|W_j)\nn\\
&&+\sum^N_{i=1}H(X^n_i|M_0)-n\epsilon'.\eqa
Together with (\ref{eq:ach2}), we get,
\beq \sum^N_{i=1}H(X^n_i|M_0)-\sum^n_{j=1}H(X_{1j}X_{2j}\cdots X_{Nj}|W_j)\leq n\epsilon^{''} .\label{eq:upper1}\eeq
where $\epsilon^{''}=(N+1)\epsilon$. On the other hand, we have,
\bqa &&\sum^N_{i=1}H(X^n_i|M_0)\nn\\
&=&\sum^N_{i=1} \sum^n_{j=1}H(X_{ij}|X^{j-1}_iM_0),\\
&\geq&\sum^N_{i=1} \sum^n_{j=1}H(X_{ij}|X^{j-1}_1,X^{j-1}_2,\cdots,X^{j-1}_N,M_0),\\
&=&\sum^N_{i=1} \sum^n_{j=1} H(X_{ij}|W_j).\label{eq:upper2}\eqa
Combine (\ref{eq:upper1}) and (\ref{eq:upper2}), we have,
\beq \sum^n_{j=1} \Big[ \sum^N_{i=1}H(X_{ij}|W_j)-H(X_{1j}X_{2j}\cdots X_{Nj}|W_j)\Big]\leq n\epsilon^{''}.\eeq
Hence, we have,
\beq \frac{1}{n}\sum^n_{j=1} \delta^{(j)}_2 \leq \epsilon^{''}.\eeq
This completes the proof of Lemma \ref{lemma2}.$\Box$

Now, from Lemma \ref{lemma2} and (\ref{eq:r0bound}), we get,
\beq R_0\geq \frac{1}{n}\log{{\cal M}_0}\geq H(X_1,X_2,\cdots,X_N)-\Gamma_1(\eta).\eeq
Together with the continuity of $\Gamma_1(\cdot)$, we have, as $n\rightarrow \infty$,
\bqa R_0&\geq&  H(X_1,X_2,\cdots,X_N)-\Gamma_1(0),\\
&=&C(X_1,X_2,\cdots,X_N).\eqa
This completes the proof of converse part.$\Box$

We now prove the achievability part, that is, let the joint
distribution $P(x_1,x_2,\cdots,x_N)$ be given, we have,
\begin{theorem}[Achievability]\label{them3}
\beq C_1\leq C(X_1,X_2,\cdots,X_N).\eeq
\end{theorem}
Our proof mainly involves generalizing Gray-Wyner source coding
network \cite{Gray_Wyner_74} to that of $N$ sources. The system
model we considered here is the same as Fig.~\ref{fig:model1}
described in section II except that definition \ref{def2} is
replaced by,
 \begin{definition}
 A rate tuple $(R_0,R_1,\cdots,R_N)$ is said to be {\em achievable} if for all $\epsilon>0$, we can find an $n$ sufficiently large such that there exists a  $(n,2^{nR_0},2^{nR_1},\cdots,2^{nR_N})$ code with
\bqa P^{(n)}_e&\leq& \epsilon .\eqa

 \end{definition}
 Our purpose is to find all achievable rate tuples $(R_0,R_1,\cdots,R_N)$.
The rate region of this source coding problem is summarized in the following theorem.
\begin{theorem}\label{theorem4}
For the source coding model described above, a rate tuple $(R_0,R_1,\cdots,R_N)$ is achievable if and only if the following conditions are satisfied,
\bqa R_0&\geq& I(X_1,X_2,\cdots,X_N; W),\label{eq:them41}\\
R_i&\geq& H(X_i|W),\label{eq:them42}\eqa
for $i=1,2,\cdots,N$, and for some $W\sim P(w|x_1,x_2,\cdots,x_N)$, where $W\in {\cal W}$ and $|{\cal W}|\leq\prod^N_{i=1}|{\cal X}_i|+2$.

\end{theorem}

{\em Proof of Theorem \ref{theorem4}} (Sketch):

For the achievability part, we want to show that for any rate tuple $(R_0,R_1,\cdots,R_N)$ that satisfies above conditions, we can construct a $(n,2^{nR_0},2^{nR_1},\cdots,2^{nR_N})$ code such that the decoding error $P^{(n)}_e\rightarrow 0$ as codeword length $n\rightarrow \infty.$

{\em Codeword Generation}: for any given distributions $P(x_1,x_2,\cdots,x_N)$ and $P(w|x_1,x_2,\cdots,x_N)$, we calculate the marginal distribution $P(w)$.
\begin{enumerate}
\item  Codebook $C_0$: we first randomly generate $2^{nR_0}$ sequences  $w^n$ i.i.d. $\sim P(w)$, and index them by $m_0\in\{1,2,\cdots,2^{nR_0}\}$.

\item Codebook $C(X_i)$: for each $i=1,2\cdots, N$, for each $x^n_i\in {\cal X}^n_i$, randomly put them into $2^{nR_i}$ bins and index them bins by $m_i\in\{1,2,\cdots, 2^{nR_i}\}$.
    \end{enumerate}

{\em Encoding}:
\begin{enumerate}
\item for each source sequences $(x^n_1,x^n_2,\cdots,x^n_N)$, encoder $f_0$ finds a $w^n(m_0)\in C_0$ such that $(x^n_1,x^n_2,\cdots,x^n_N,w^n(m_0))\in T^n_{\epsilon}$, where $T^n_{\epsilon}$ is the jointly typical set as defined in \cite{Cover:Book91}, and send the index $m_0$ to the decoder.  If there is no more than one $w^n$, choose the sequence $w^n$ with the smallest index; if there exist no such sequence, choose sequence $w^n(1)$,

\item for $i=1,2,\cdots,N$, encoder $f_i$ sends the bin index $m_i$ of sequence $x^n_i$.

\end{enumerate}

{\em Decoding}: for $i=1,2,\cdots,N$, decoder $i$ looks at bin $m_i$ for codebook $C(X_i)$ and finds the sequence $\hat{x}^n_i$ such that $(\hat{x}^n_i,w^n(m_0))\in T^n_{\epsilon}$. If there is more than one or none such sequence, declare an error.

{\em Error analysis}: Assuming $m_i$, $i=0,1,\cdots,N$ are the chosen indices for encoding $(x^n_1,x^n_2,\cdots,x^n_N)$. There are three error events.
\begin{enumerate}
\item $E_{1}$:  $(x^n_1,x^n_2,\cdots,x^n_N,w^n(m_0))\notin T^n_{\epsilon}$ for all $m_0\in \{1,2,\cdots,2^{nR_0}\}$.
    \item $E_2$: $(x^n_i,w^n(m_0))\notin T^n_{\epsilon}$ for each $i$.
    \item $E_3$: for some $i$, there exists $\tilde{x}^n_i\neq x^n_i$ in bin $m_i$ of codebook $C(X_i)$ such that $(\tilde{x}^n_i,w^n(m_0))\in T^n_{\epsilon}$.
\end{enumerate}
Hence, \beq P^{(n)}_e\leq P(E_1)+P(E_2|E^c_1)+P(E_3|E^c_1,E^c_2).\eeq
By some standard argument, we can get, as $n\rightarrow \infty$,
\begin{enumerate}
\item $P(E_1)\rightarrow 0$ if
\beq R_0\geq I(X_1,X_2,\cdots,X_N; W)+\epsilon,\eeq
\item $P(E_2|E^c_1)\rightarrow 0$,
\item $P(E_3|E^c_1,E^c_2)\rightarrow 0$ if  for each $i=1,2,\cdots,N$,
\beq R_i\geq H(X_i|W)+\epsilon.\eeq

\end{enumerate}

This completes the achievability proof.

For the converse part, we want to show that for any achievable rate tuple $(R_0,R_1,\cdots,R_N)$, it should satisfy
(\ref{eq:them41}) and (\ref{eq:them42}).

By Fano's inequality, we have
\beq H(X^n_i|M_iM_0)\leq n\epsilon.\eeq
Hence, we have, for $i=1,2,\cdots, N$
\bqa &&nR_i\nn\\
&\geq&H(M_i),\\
&\geq&H(M_i|M_0),\\
&\geq&H(M_i|M_0)+H(X^n_i|M_iM_0)-n\epsilon,\\
&=&H(X^n_iM_i|M_0)-n\epsilon,\\
&=&H(X^n_i|M_0)-n\epsilon,\\
&=&\sum^n_{j=1}H(X_{ij}|M_0X^{j-1}_i)-n\epsilon,\\
&\geq&\sum^n_{j=1}H(X_{ij}|M_0,X^{j-1}_1,X^{j-1}_2,\cdots,X^{j-1}_N)-n\epsilon.\eqa
and
\bqa &&nR_0\nn\\
&\geq&H(M_0),\\
&\geq&I(M_0; X^n_1,X^n_2,\cdots,X^n_N),\\
&=&\sum^n_{j=1}I(M_0;X_{1j}X_{2j}\cdots X_{Nj}|X^{j-1}_1X^{j-1}_2\cdots X^{j-1}_N),\\
&=&\sum^n_{j=1}I(M_0X^{j-1}_1X^{j-1}_2\cdots X^{j-1}_N;X_{1j}X_{2j}\cdots X_{Nj}).\eqa
Define $W_j=(M_0,X^{j-1}_1,X^{j-1}_2,\cdots, X^{j-1}_N)$, and using a standard time sharing argument, we can get, for $i=1,2,\cdots,N$,
\bqa R_i&\geq& H(X_i|W)-\epsilon,\\
R_0&\geq& I(X_1X_2\cdots X_N; W).\eqa
Let $n\rightarrow \infty$, then $\epsilon\rightarrow 0$, and this completes the proof of converse. The cardinality bound can be obtained using the technique introduced in \cite[Appdendix C]{ElGamal&Kim:Lecture}. We skip the details. This completes the proof of Theorem \ref{theorem4}.$\Box$

Now we proceed to prove Theorem \ref{them3}. We will show that if $R_0>C(X_1,X_2,\cdots,X_N)$, it is achievable for Model I.

Let $R_0>C(X_1,X_2,\cdots,X_N)$ and any $\epsilon>0$ be given and let random variables $(X_1,X_2,\cdots,X_N,W)$ satisfy (\ref{theorem1.1}) and (\ref{theorem1.2}), such that
\beq C(X_1,X_2,\cdots,X_N)=I(X_1X_2\cdots X_N;W).\eeq
Notice that, the existence of such random variables is guaranteed by Lemma  \ref{lemmawnyer}. Now define
\beq \epsilon_1=\min \{\frac{\epsilon}{N+1}, R_0-C(X_1,X_2,\cdots,X_N)\},\eeq
and hence $\epsilon_1>0$. By Theorem \ref{theorem4}, there exists a $(n,{\cal M}_0,{\cal M}_1,\cdots,{\cal M}_N)$ code with $P^{(n)}_e\leq \epsilon'$ and $\epsilon' \leq \epsilon_1$. Hence,
\bqa \frac{1}{n}\log{{\cal M}_0}&\leq& C(X_1,X_2,\cdots,X_N)+\epsilon_1\leq R_0,\\
\frac{1}{n}\log{{\cal M}_i}&\leq& H(X_i|W)+\epsilon_1.\eqa
Hence, we have,
\bqa&& \sum^N_{i=0}\frac{1}{n}\log{\cal M}_i\nn\\
&\leq&C(X_1,X_2,\cdots,X_N)+\sum^N_{i=1}H(X_i|W)+\epsilon,\\
&\stackrel{(a)}{=}&H(X_1,X_2,\cdots,X_N)+\epsilon.\eqa
where $(a)$ is from condition (\ref{theorem1.2}). Thus, condition (\ref{eq:ach2}) is also satisfied. This implies that $R_0$ is achievable in Model I, which completes the proof of achievability part. This completes the proof of Theorem \ref{them3}.$\Box$

\subsection{Proof of $C_2=C_{X_1,X_2,\cdots,X_N}$.}
In this section, we prove the second part of theorem 1, that is  $C_2=C(X_1,X_2,\cdots,X_N).$  We have the following theorem.
\begin{theorem}
\bqa C_2\geq C(X_1,X_2,\cdots,X_N),\label{eq:conv}\\
C_2\leq C(X_1,X_2,\cdots,X_N).\label{eq:achc2}\eqa

\end{theorem}

For the converse part , that is (\ref{eq:conv}), the proof follows
almost the same line as in \cite[Section 5.2]{Wyner_CI_75IT}. For
the achievability part, that is (\ref{eq:achc2}), the proof
follows similarly as in \cite[Seciton 6.2]{Wyner_CI_75IT} by
applying ${\cal U}={\cal X}_1\times{\cal X}_2, \cdots\times{\cal
X}_N$ in \cite[Theorem 6.3]{Wyner_CI_75IT}. We omit the details
here.

\bibliographystyle{IEEEbib}

\end{document}